\documentclass[10pt,journal]{IEEEtran}
\usepackage{amsmath,amssymb,amsfonts}
\usepackage{algorithmic}
\usepackage{algorithm}
\usepackage{graphicx}
\usepackage{mathtools}
\usepackage{xfrac}
\usepackage{textcomp}
\usepackage{booktabs}
\usepackage{amssymb}
\usepackage{subfigure}
\usepackage{array}
\usepackage{float}
\usepackage{stfloats}
\usepackage{authblk}
\usepackage{xcolor}
\usepackage{cite}
\usepackage{subfiles}
\usepackage{listings}
\usepackage{balance}
\usepackage{xspace} 
\usepackage{subfigure}
\usepackage[english]{babel}
\usepackage{amsthm}
\usepackage{url}
\usepackage{hyperref}
\ifCLASSINFOpdf
 
\else
 
\fi
\theoremstyle{definition}

\begin{document}

\title{LSAI: A Large Small AI Model Codesign Framework for Agentic Robot Scenarios}

\author{Longyu Zhou, Supeng Leng, Tianhao Liang, and Jianping Yao
\thanks {L. Zhou is with the Information Systems Technology and Design Pillar, Singapore University of Technology and Design, 487372, Singapore (e-mail: zhoulyfuture@outlook.com) (Corresponding author is Jianping Yao).
}
\thanks{S. Leng is with the School of Information and Communication Engineering, University of Electronic Science and Technology of China, Chengdu 611731, China (e-mail:spleng@uestc.edu.cn).
}
\thanks{T. Liang is with the School of Information Science and Technology, Harbin Institute of Technology (Shenzhen), Shenzhen 518055, China (E-mail: liangth@hit.edu.cn).}
\thanks{J. Yao is with the School of Information Engineering, Guangdong University of Technology, Guangzhou 510006, China, and also with Guangdong Provincial Key Laboratory of Future Networks of Intelligence, The Chinese University of Hong Kong at Shenzhen, Shenzhen 518172, China (e-mail: yaojp@gdut.edu.cn).
}
}
\maketitle

\begin{abstract}
\label{sec_abstract}
The development of Artificial Intelligence (AI) has enabled agentic robots an appealing paradigm for various applications, such as research and rescue in complex environment. In this context, the next wireless communication technology facilitates robot cooperation for efficient environment sensing and exploration. However, traditional AI solutions cannot always provide reasonable resource utilization decisions, which makes it challenging to achieve both accurate and low-latency research and rescue. To address this issue, we propose a, LSAI, a large small AI model codesign framework to achieve highly accurate and real-time robot cooperation with deep interaction between large AI model and small AI model. We first propose an attention-based model aggregation for LAI construction. It can assist agentic robots in accurately sensing physical environments. Next, we design an adaptive model splitting and update algorithm to enable the robots to perform accurate path planning for high-efficiency environment sensing with low energy consumption. Finally, we demonstrate the effectiveness of our proposed LSAI framework. The simulation results indicate that our solution achieves sensing accuracy of up to 20.4\% while reducing sensing cooperation latency by an average of 17.9\% compared to traditional AI solutions.

\end{abstract}

\begin{IEEEkeywords}
large AI model, small AI model, robot, model splitting, model update.
\end{IEEEkeywords}
\section{Introduction}
\label{sec:introduction}
\IEEEPARstart{W}{ith} the rapid progress of artificial intelligence (AI) and sixth-generation (6G) mobile networks, robots are emerging as a key platform for future intelligent applications~\cite{AI_6G,AI_driven}. Their compact size and operational flexibility make them suitable for a wide range of tasks. In large-scale scenarios, such as emergency response, security monitoring, and healthcare, robots can provide accurate sensing and scene assessment. With AI support, robots can also adapt to complex environments, evaluate tasks autonomously, and make intelligent decisions, which improves mission effectiveness. In a typical search-and-rescue scenario, robots first perform collaborative sensing over multiple areas and then upload sensed data to a central server. The server analyzes the collected information and generates cooperative path-planning decisions. Based on these decisions and their local states, robots further adjust their sensing strategies to enable accurate search and timely rescue.

Despite these advantages, efficient multi-robot collaboration remains challenging. First, robots have limited onboard sensing, computing, and communication resources. These limitations reduce their ability to collect information in real time, process heterogeneous data efficiently, and exchange messages reliably. As a result, mission efficiency decreases and the risk of collision may increase~\cite{collision}. Second, search-and-rescue environments are often highly dynamic and complex. Under such conditions, limited computing capability makes it difficult to generate proper search paths, which increases search time and reduces coordination quality. In addition, harsh physical environments may severely degrade wireless communication quality, further weakening inter-robot cooperation~\cite{cooperative_data}. Third, search tasks are usually random, distributed, and delay-sensitive. Existing server-centric centralized architectures rely on long-distance data transmission and remote decision making. This introduces high latency and prevents robots from obtaining timely path-planning updates, which is unfavorable for time-critical missions~\cite{real-time_decision}. Therefore, a more efficient and environment-adaptive collaborative framework is needed for robotic search services.

Edge computing provides a promising way to reduce transmission latency by moving computing resources closer to robots~\cite{edge_}. This shortens communication distance and improves the timeliness of decision support. However, the randomness of search tasks and the strong dynamics of practical environments place significant pressure on resource-limited edge servers. Large AI models (LAMs) offer strong capabilities in representation learning, task analysis, and decision support~\cite{geng}. By training on massive sensing data, they can capture task distributions and environmental patterns, and they can be further fine-tuned for specific search objectives and latency requirements. In this way, LAMs can generate accurate environment understanding and high-quality cooperative path-planning decisions. Nevertheless, directly deploying only LAMs at the edge is still insufficient. The inference delay of LAMs may become excessive when processing complex heterogeneous sensing data. Their update process is also costly under rapidly changing environments. These limitations make it difficult to guarantee real-time decision output~\cite{sensing_limitation}. This motivates the need for a new collaborative intelligence framework that jointly considers communication, computation, sensing, and path planning.

To address the above issues, this paper proposes a, LSAI, large and small AI model codesign framework for agentic robotic search scenarios. In the proposed framework, robots use multimodal sensors to collect heterogeneous information and train local small models with onboard resources. The lightweight parameters of these small models are uploaded to the edge server, where they are fused to construct an edge-side large model. The large model is used to estimate complex environments and predict environmental changes more accurately. It also generates proper sensing path planning decisions by jointly considering the energy, location, and velocity of robots. At the same time, robots use updated local small models to analyze local environmental variations and refine sensing paths online. The main contributions of this work are threefold. 

\begin{itemize}
    \item We develop a, LSAI, a large–small model codesign framework for agentic robotic scenarios. The proposed framework integrates terminal-edge computing resources and enhances collaborative sensing in complex environments. It also improves sensing accuracy and supports scenario-adaptive multi-task cooperation, which provides an architectural basis for future multi-robot intelligent applications.
    \item We design an attention-based customized model fusion algorithm to support efficient LAI model construction. The proposed algorithm selects suitable SAI parameters according to the relative positions of robots and their historical evaluation records, and then performs real-time parameter fusion to build an effective large model for accurate environment estimation. Based on this model, the system can further plan sensing paths according to robot energy states, which supports high-efficiency collaborative sensing. 
    \item We develop an adaptive model splitting algorithm to improve SAI adaptability. This algorithm extracts suitable parameters from the large model according to robot mobility trajectories and uses them to update SAI models. As a result, robots can better capture local environmental dynamics and adjust sensing paths in an energy-efficient manner. We also present a search-and-rescue use case and provide both qualitative and quantitative evaluations. We also discuss the future opportunities of large–small model collaboration.
\end{itemize}

In this article, we first describe our proposed framework. We then discuss our proposed algorithms to support our framework with convincing results. Finally, we give future research directions for our solution.

\section{Large Small AI Model Codesign Framework for Agentic Robot Scenarios}
\label{sec_system_model}
In this section, we provide clear descriptions for our proposed Large Small AI Model (LSAI) Codesign framework.

\begin{figure*}[t]
\centerline{\includegraphics[width=\linewidth]{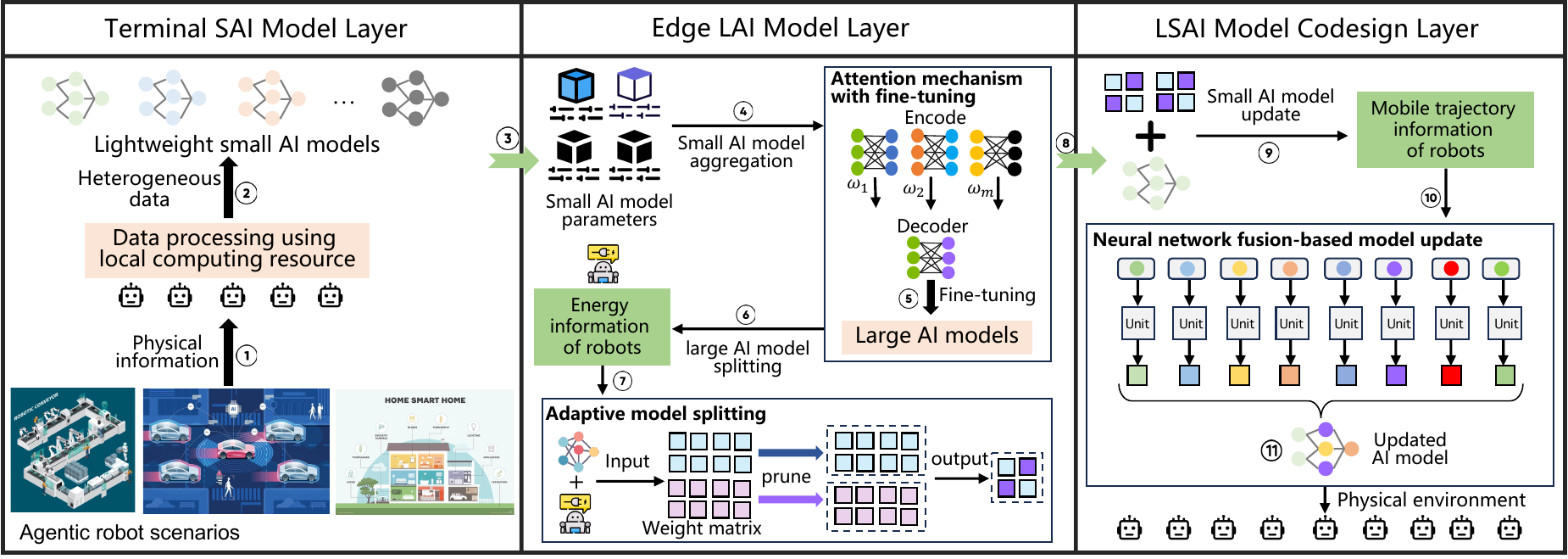}}
\caption{Illustration of large small AI model (LSAI) codesign framework for robots-based research and secure scenarios with three layers: the terminal small AI model layer, the edge large AI model layer, and the LSAI model codesign layer.}
\label{fig_system_model}
\end{figure*}

As shown in Fig.~\ref{fig_system_model}, our LSAI framework can be widely applied to multiple robots-based mobile scenarios with cross-layer computing resource cooperation for high-efficiency scenario estimation and prediction. In such a scenario, we can deploy multiple robots with strong computing capability as edge robots to manage different areas. For each area, we can deploy multiple terminal robots to implement relevant missions. In this context, edge robots can firstly enable robots to perform cooperative environment sensing to construct Small AI (SAI) models based on local sensing information. These SAI model parameters are transmitted to edge robots to construct a global Large AI (LAI) model. The LAI model can provide accurate environment estimation for high-efficiency cooperative sensing for robots based on their attribute information (such as positions, velocities, and trajectories). In addition, the LAI model can provide suitable model parameters to update SAI models for improving decision-making accuracy based on local environment information of robots. In this case, we can collaborate LAI with SAI to ensure real-time and accurate mission implementation in different robots-based mobile scenarios. For clarity, we categorize our framework into three distinct layers: the terminal SAI model Layer, the edge LAI model Layer, and the LSAI model codesign Layer.   

\subsection{Terminal SAI Model Layer}
We take the search and rescue as an instance. The terminal SAI model layer performs a fundamental component with two main functions: cooperative data collection and SAI model construction. Specifically, robots can autonomously collect their local environmental information using onboard sensors, such as camera, ultrasonic, and UltraWide Band (UWB) sensors. The sensing information includes the information of topography and terrain and the status of neighboring robots. These robots can use local computing resource to process heterogeneous information. The processing results are used to construct local SAI models for high-efficiency sensing cooperation by planning suitable sensing paths. However, these path planning decisions might be unreasonable only based on local physical information. To address this issue, we can allow robots to implement cooperative model training through an information exchange operation among the robots to ensure high training accuracy.

In details, to ensure high-efficiency data collection, we invoke a dynamic sensing scheme~\cite{sensing_coverage}. It can improve the utilization of sensing resources of robots to ensure comprehensive data collection. The method can assist robots in planning feasible sensing paths to efficiently reduce blind sensing zones. Based the potential method, we then use the Jaccard index to control the overlap between any two robots for high-efficiency cooperative sensing. The sensing result can strongly support accurate SAI model training based on a reinforcement learning architecture~\cite{Convention_DDPG}. During the training processing, we can constructively implement a sharing operation of training estimation results among robots to improve the training accuracy for suitable sensing path planning and the reduction of physical collision. 

\subsection{Edge LAI Model Layer}
%model aggregation and model splitting一种基于注意力机制的定制化模型融合算法，该算法能够基于机器人的相对位置和历史评估经验，提取合适的小模型参数去执行实时的参数融合.....自适应模型裂解算法，该算法能够基于机器人的移动轨迹信息，从大模型中提取合适的模型参数，用于更新本地机器人的小模型，
We mainly empower two operations for accurate environment estimation and sensing path planning in the edge LAI model layer: the SAI model aggregation and adaptive LAI model splitting. The former is implemented to perform accurate environment estimation. The latter is enabled to optimize the SAI models for accurate path planning for agentic robots. For the SAI model aggregation, we allow the edge robots to conduct SAI model aggregation process using our proposed attention-based model aggregation method. This method ensures the aggregation of a high-accuracy LAI through selection of feasible model parameters. To ensure high computing resource utilization, we can reasonably collaborate computing resources of agentic robots and edge robots to implement accurate model splitting based on mobile trajectory information of robots. Consequently, we can achieve a high-accuracy LAI model construction that performs accurate environment sensing and estimation in robot-based mobile scenarios.

Based on this, we implement an adaptive model splitting action based on the LAI to ensure accurate sensing path planning. Specifically, we propose an adaptive model splitting algorithm. The algorithm can assist robots in dynamically splitting the LAI to exact feasible model parameters based on current mobile trajectories of robots (details in Sec~\ref{sec_algorithm}). The model parameters can assist the current SAI models in adjusting the connection of neural units for high-accuracy path planning exploration. In addition, the agentic robots can implement model parameter exchanges to further optimize the cooperative sensing path decision with physical collision avoidance. Furthermore, the spitted models can effectively assist robots in improve the target identification accuracy based on the analysis of environment estimation results. Our approach ensures accurate sensing path planning and sensing target identification through the real-time information exchanges among robots with feasible model parameters.

In this context, we achieve three functions in the edge LAI model layer: communication resource scheduling, cooperative path planning, and accurate target identification. Specifically, the LAI can schedule feasible communication resources to collaborate with multiple edge robots to implement cooperative environment estimation. The estimation results can be used to assist agentic robots to perform feasible sensing path planning. With the mobile trajectory information of each robots, the edge robots can implement high-efficiency model splitting to extract feasible model parameters for optimization of SAI models. The optimization way can further improve the path planning efficiency significantly. Subsequently, we can utilize the estimation results and path planning decisions to high-efficiency implement target identification for real-time mission implementation. This accurate identification results can also assist robots in optimizing the communication resource scheduling in return for saving communication overhead. The close-loop design can obviously ensure high-efficiency robot cooperation for complex mobile scenarios with computing and communication resource saving.

\subsection{LSAI Model Codesign Layer}
%adaptive model update
The LSAI model cooperation is designed to serve the SAI model update based on information of robots and environment, including positions, velocities, energy and obstacles. It is implemented on multiple agentic robots via information exchanges. Explicitly, we propose an neural network fusion-based model update algorithm. First, the LAI splitting method can extract feasible model parameters based on robots' mobile trajectories. Our algorithm combines the feasible model parameters to the current SAI models to optimize the SAI model performance for real-time and accurate path adjustment. Furthermore, our algorithm further ensure soft sensing cooperation by enabling information sharing among agentic robots. This assists agentic robots in optimizing cooperative sensing decisions through path adjustments to achieve collision-free and low-latency cooperative sensing in complex research scenarios. We provide detailed descriptions of our algorithm.
%environment sensing(robot)-->environment estimation(aggregation)(LAI)-->robot path planning(splitting)(LAI)-->robot path adjstment(update)(robot)--->accurate environment sensing (robots).

Explicitly, we first implement a feature alignment operation for extracted and current AI model parameters to ensure smooth model combination. Then, we implement model fusion operation at the stage of model input which can achieve cross-module learning for enhancing model accuracy. Based on this, we can formulate the feasible model training function to train the fusion model. To ensure high-efficiency training performance for accurate path planning decisions, we can implement the information exchanges that allows for cooperative training with a high computing resource utilization. Meanwhile, this way can also ensure low communication overhead by transmitting only lightweight training model parameters. In this context, we can achieve persistent path planning performance through enhanced robot collaboration.

On the other hand, our model update method enables agentic robots to perform dynamic path optimization for accurate environment sensing by predicting the trajectories of neighboring robots based on their positions and environment information. the improvement of environment sensing is reflected in two key aspects: cooperative path adjustment and autonomous model optimization. Specifically, robots can engage in decision-sharing behavior among neighbors to obtain sensing path decisions. When overlapping sensing paths are detected, agentic robots can utilize the our algorithm to update a new path. Meanwhile, the new path decision is shared to conduct other agentic robots to optimize sensing paths for cooperative sensing convergence. However, robots cannot always perform accurate path adjustment due to time-varying sensing environment. To address this issue, the robots can implement dynamic model optimization using our proposed algorithm. Consequently, our approach jointly considers potential physical collisions and changes in environment to optimize multiple paths for high-accuracy environment sensing.

\section{Algorithm Designs and Performance Evaluation}
\label{sec_algorithm}
In this section, we provide a detailed description of our solution, which consists of two parts: \emph{attention-based model aggregation for LAI construction} and \emph{adaptive LAI splitting and SAI model update}. Based on our algorithm, we then give the evaluation results to verify the effectiveness of our solution.

\subsection{Attention-based Model Aggregation for LAI Construction}
\label{sec_sensing}
\begin{figure*}[t]
\centerline{\includegraphics[width=.75\linewidth]{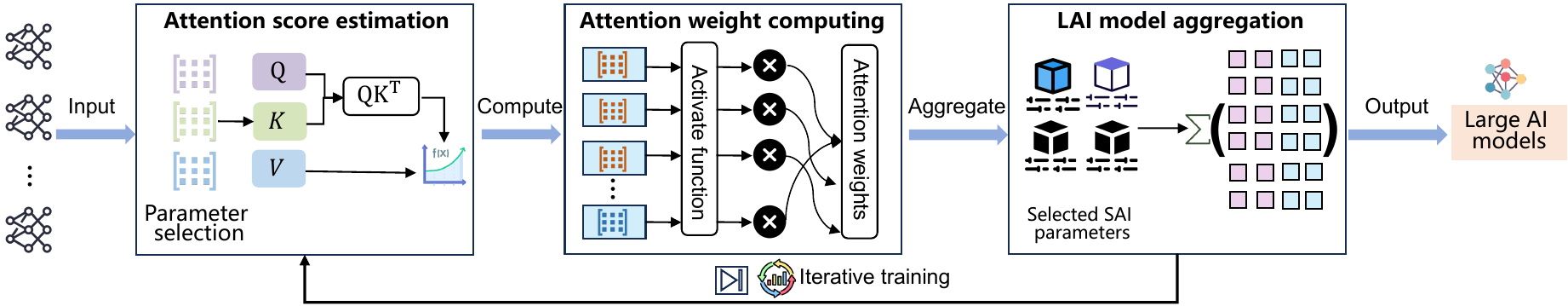}}
\caption{Illustration of model aggregation for LAI construction. }
\label{fig:LAI_construction}
\end{figure*}

We can use the popular Federated Averaging method~\cite{FAM} to implement model aggregation for LAI construction. However, it might cause a low training convergence with diverse types of sensing data. To address the issue, we invoke a attention mechanism to dynamically optimize weights of model parameters for rapid training with suitable parameter selection. We present the implementation process of LAI construction illustrated in Fig.~\ref{fig:LAI_construction}. Compared with traditional aggregation methods such as simple averaging~\cite{Aggregation}, our approach allows the edge robots to adaptively emphasize more informative model updates while reducing the influence of unreliable ones. The specific implementation process mainly includes four steps: \textit{collection of local SAI models}, \textit{attention score estimation}, \textit{attention weight computing}, \textit{LAI model aggregation}, \textit{iterative model training}.

Before that, we can use an mature Deep Determined Policy Gradient (DDPG) architecture to implement SAI model training~\cite{Convention_DDPG}. With the SAI models, the edge robots can receive model parameters from all involved agentic robots. At the stage of the collection of local SAI models, the edge robot possesses multiple locally trained models, each representing the learning outcome from a different robot’s sensing information. The edge server can simply average these models to obtain the next global model. However, because different robots may have significantly different data distributions, treating all robots' models equally may lead to suboptimal updates. To address this limitation, we design attention-based model aggregation method as an additional embed to estimate the relative importance of each model before performing aggregation.

Specifically, we can enable the edge robot to evaluate the relevance or contribution of each uploaded model using an attention scoring mechanism at the stage of attention score estimation. The purpose of this step is to measure how well each model aligns with the global optimization objective. To perform this evaluation, the edge robot compares each model with the current global model. The comparison reflects whether the update direction provided by the robot is consistent with the general learning direction of the global model. If a model shows a high degree of consistency with the global model, it is considered more reliable and obtain a higher attention score.Conversely, if the update deviates significantly from the global model, its contribution may be less beneficial and therefore receives a lower score.

Based on the estimation scores, the edge robot converts these scores into normalized aggregation weights at the stage of attention weight computing. The normalization process ensures that the weights are comparable across robots and that the sum of all weights equals one. Each robot model is then assigned a corresponding attention weight. Models with higher weights will contribute more significantly to the global update, while models with lower weights will have a smaller impact on the aggregation result. This adaptive weighting mechanism enables the training robot system to prioritize high-quality model updates while mitigating the negative influence of noisy or inconsistent updates.

With the computing results, the edge server aggregates these models using a weighted combination strategy. Specifically, each model is multiplied by its corresponding attention weight, and the weighted models are then combined to produce a new global model. This weighted aggregation allows the edge robot to incorporate useful information from multiple robots while giving greater emphasis to the most informative updates. As a result, the global model can evolve more effectively than with simple averaging methods.

To ensure accurate LAI construction, we finally enable an iterative training mode consisting of multiple communication rounds. In each round, a subset of robot performs local training using DDPG architecture and sends updated models to the edge robot. The server evaluates the relative importance of these models using the attention mechanism and aggregates them accordingly. Over successive iterations, the global model progressively improves as it incorporates information from diverse robots' data. The attention-based aggregation strategy ensure that the most informative updates contribute more strongly to the learning process, thereby enhancing model robustness and overall performance in complex search and secure scenarios.

\subsection{Adaptive LAI Splitting and SAI Update}
\label{sec_model_transfer}

\begin{figure*}[t]
\centerline{\includegraphics[width=.75\linewidth]{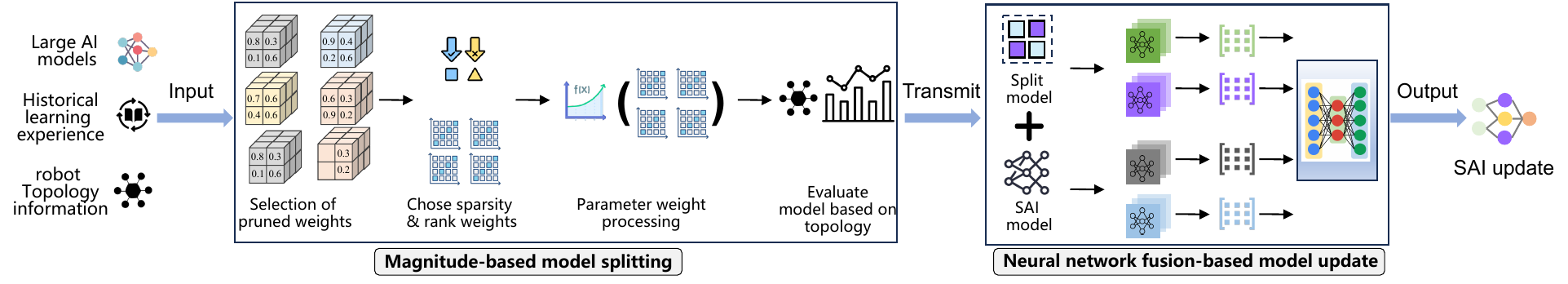}}
\caption{Illustration of LAI splitting and SAI update. }
\label{fig:split_update}
\end{figure*}

%teams usually exclude a few sensitive parts at first, such as embeddings, normalization layers, or very small layers, and focus on the main weight tensors where most parameters live. At the same time, they choose a pruning scope, either global across the whole model or layer-wise within each layer, because that decision affects how aggressively different parts of the model will be reduced.
However, the LAI model cannot conduct robots to implement real-time environment estimation and path adjustment due to high dynamics of physical environment, we design a magnitude-based weight LAI splitting method to provide feasible model parameters to robots for high-accuracy environment estimation. In addition, we then propose a neural network fusion-based model update method to implement SAI update for real-time path adjustment and optimization with energy saving and collision avoidance. We give specific implementation process in Fig.~\ref{fig:split_update}.

In details, for the LAI splitting operation, we first set a LAI model sparsity level. Once that level is chosen, we implement model scanning with trained weights and ranks them by their magnitude. The weights with the smallest magnitudes are treated as the least important and selected for removal. In standard implementations, this selection is turned into a mask that marks which weights stay active and which ones are pruned. Based on the mask, we can sparse the model through setting the selected low-magnitude weights to zero. In this case, we divided the trained weights into two groups, the important weights and the low-importance weights. 

Based on this, we fine-tune the model so the selected model weights can adapt and recover lost accuracy. we implement a retraining step to update the surviving parameters while the pruned weights remain inactive. With an iteration operation, the temporary pruning reparameterization is removed, the sparse model is saved, and the final model is evaluated for accurate environment estimation based on the robot topology information. In this context, we can obtain the customized model splitting results for each robot.

With the pruned LAI sub-model, we can implement a model update operation through a neural network fusion method. Firstly, we can qualify the SAI model $\text{SAI}_{i}$ of robot $i$ as a graph $G$. In this case, we can combine $\text{SAI}_{i}$ and LAI sub-model $\text{Sub}_{i}$ by exploring a feasible branch in $G$. Take SAI model neuron $v$ at the r-th branch as an instance, the model combination can be formulated as where $f_v^r(*)$ is a graph operation for model combination with neuron aggregation; $g^r$ is a learnable dimension transformation function for model combination. We use the method to explore the optimal combination branch according to the objective of minimizing energy consumption while ensuring collision avoidance. We can obtain the updated $\text{SAI}_{i}^{u}$ using $f_v^r(*)$ and $g^r$. It is noted that when a large number of robots involve, edge robots cannot always implement the real-time model update with vast of model parameters. In this case, edge robots can inform terminal agentic robots to implement re-training operation based on fresh local information. Specifically, we can acquire SAI models $\{ \text{SAI}_1, \cdots, \text{SAI}_i \}$ using our invoked DDPG architecture with low numbers of iterations. The SAI models are transmitted to corresponding edge robots which can select feasible model parameters to distribute to the agentic robots based on the analysis of SAI model parameters. In this case, the agentic robots can perform real-time model update for accurate path optimization. However, the model update can lead to a performance decline when processing vast cross-regional robots with a high latency. In this case, we allow edge robots to run a re-training operation for high-accuracy environment estimation. Based on this, we can obtain feasible sensing decisions with the aid of large and small AI models to serve mobile robot scenarios.

\subsection{Results}
\label{sec_results}
We build a 3 km $\times$ 3 km robot-based research and rescue scenario with multiple robots and detected targets using Gazebo on the NVIDIA GB300 NVL72 server. To perform high-efficiency and immersive simulation, we use the differential drive kinematics theory to plan the sensing paths of all the robots for data collection and environment estimation~\cite{differential} using diverse onboard sensors including cameras, LiDAR, and the Inertial Measurement Unit (IMU). The robots can obtain the position information using the Global Positioning System (GPS) sensor. The energy information is obtained via voltage sensors. For the sensing implementation, we can train to obtain the SAI models based on DDPG architecture using the \href{https://www.kaggle.com/datasets/lastman0800/warehouse-manageent-using-ai}{intelligent robots dataset}. We implement the SAI model aggregation operation using our proposed attention mechanism. The model update operation is enabled to obtain the optimized SAI models using our neural network fusion based model update method. Our method consists of three hidden layers, and each hidden layer has 64 neural units. The learning rate and batch size are set to 0.99 and 128, respectively. We use a sigmoid activation function to implement data training. Considering the random deployment of detected targets, we adopt a stochastic gradient descent optimization method to explore feasible path planning and optimization through an iterative learning process. In addition, the \href{https://www.kaggle.com/datasets/ziya07/robot-interaction-and-network-performance-dataset}{robot interaction dataset} is used to achieve robot cooperation for cooperative sensing with key interaction variables: robot ID, type of parcels, sensing start and end times, robot communication types, and data packet size. Three primary metrics are evaluated:

\begin{enumerate}

\item \emph{Sensing accuracy}: This metric reflects accuracies of model training and update for accurate sensing. It is formulated as a ratio between the number of sensed targets and the all the number of targets.

\item \emph{Path planning efficiency}: This metric estimates the performance of robot cooperation by analyzing environment data with varying numbers of robots and targets.

\item \emph{System response time}: We leverage the metric to evaluate the robot cooperation efficiency.% for real-time parcel sorting and handling.

\end{enumerate}

We provide two state-of-the-art benchmarks for comparison: a centralized large AI model implementation method~\cite{centralized_} based on a cloud computing pattern and a distributed small AI model implementation method~\cite{distributed2} with cooperative AI model construction among robots. The moving velocities of the robots are set at [5 km/h, 6 km/h] with a total of 60 robots. The number of targets is limited to [30, 50] with diverse types and positions. 

\begin{figure*}[t]
\centering
\subfigure[Sensing accuracy vs. number of robots.]{
\begin{minipage}[t]{0.3\linewidth}
\centering
\includegraphics[width=2.2in]{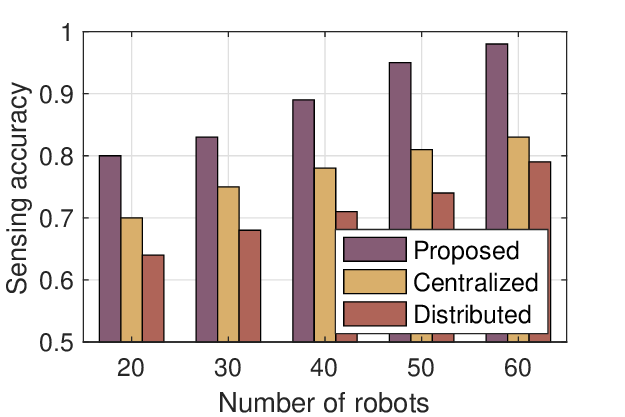}
\label{sorting}
\end{minipage}
}
\hspace{1mm}
\subfigure[Path planning efficiency. vs number of robots.]{
\begin{minipage}[t]{0.3\linewidth}
\centering
\includegraphics[width=2.2in]{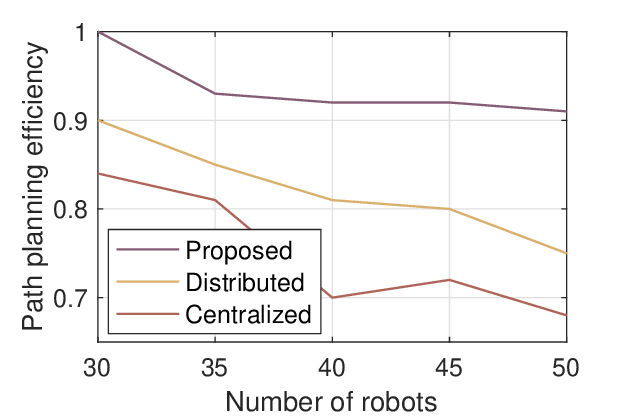}
\label{path}
\end{minipage}
}
\hspace{1mm}
\subfigure[System latency. vs. number of robots.]{
\begin{minipage}[t]{0.3\linewidth}
\centering
\includegraphics[width=2.2in]{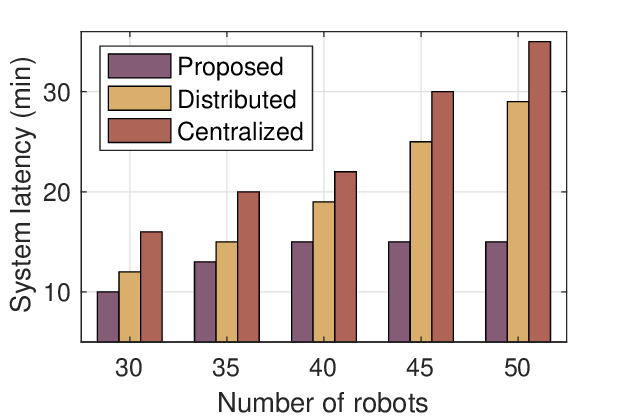}
\label{latency}
\end{minipage}
}
\caption{Performance evaluation.}
\end{figure*}

We compare parcel sorting accuracy in Fig.~\ref{sorting} under different numbers of robots. The experimental results show that sensing accuracy increases monotonically with the number of robots for all compared methods, confirming the benefit of collaborative sensing in multi-robot systems. Notably, the proposed LSAI approach consistently outperforms both the centralized large-model-only baseline and the distributed small-model-only baseline across all robot scales. Moreover, the performance gap widens as the number of robots grows, highlighting the superior scalability of the proposed LSAI framework. This advantage can be attributed to its ability to combine the high-level semantic reasoning capability of large models with the deployment efficiency and local responsiveness of small models. Therefore, our LSAI method achieves a more favorable balance between perception quality and system scalability, making it particularly suitable for large-scale multi-robot sensing tasks.

With the sensing results, we evaluate the efficiency of path planning in Fig.~\ref{path}. We can see that the proposed LSAI method consistently achieves the highest efficiency across all settings and also exhibits the smallest performance degradation, decreasing only from approximately 1.00 to 0.91, whereas the distributed method declines from about 0.90 to 0.75 and the centralized method drops from about 0.84 to 0.68. On average, the proposed method attains a path planning efficiency of about 0.936, compared with 0.822 for the distributed scheme and 0.746 for the centralized scheme. In addition, the proposed method provides an absolute improvement of approximately 0.08-0.16 over the distributed baseline and 0.13-0.23 over the centralized baseline, with the advantage becoming more evident as the number of robots increases. These results demonstrate that the proposed large-small model collaborative framework can better balance global planning capability and distributed execution efficiency, thereby maintaining superior robustness, scalability, and overall path planning performance in large-scale multi-robot systems.

Finally, we compare system implementation latency under the different numbers of robots shown in Fig.~\ref{latency}. We can see that the proposed LSAI method consistently achieves the lowest latency across all robot scales, demonstrating superior efficiency and scalability. Based on approximate values from the figure, the latency of the proposed method increases only from about 10 min at 30 robots to 15 min at 50 robots, whereas the distributed method rises from about 12 min to 29 min and the centralized method grows more sharply from about 16 min to 35 min. In terms of average performance, the proposed method yields an average latency of approximately 13 min, compared with about 20.2 min for the distributed scheme and 24.6 min for the centralized scheme. Moreover, the proposed approach reduces latency by roughly 2-14 min relative to the distributed baseline and by 6-20 min relative to the centralized baseline, with the gap becoming increasingly pronounced as the number of robots grows. These results demonstrate that the proposed large-small model collaborative framework can more effectively control system overhead while preserving coordination efficiency, thereby offering better robustness and scalability for large-scale multi-robot deployment.

\section{Future Work and Conclusion}
\label{Sec_discussion}
This section outlines future research directions for our LSAI framework in robot-based mobile scenarios.

Our solution enhances the cooperation capability between large AI model and small AI model in robot-based mobile scenarios. We can apply our LASI framework to make collaboration more adaptive, efficient, and trustworthy. A key direction is dynamic role allocation, where large models handle high-level reasoning and task decomposition. The small models perform fast local inference, control, and continuous sensing at the edge. This can be extended through online orchestration mechanisms that decide, in real time, when to invoke a large model and when a small model is sufficient, thereby reducing latency, bandwidth usage, and energy cost.

We also apply our solution to serve intelligent healthcare and wearable monitoring. Large models can integrate multimodal clinical records, medical knowledge, and long-horizon patient history to support diagnosis, treatment planning, and anomaly interpretation, while small models running on wearables, smartphones, or bedside devices continuously monitor vital signals and perform privacy-preserving local inference. This cooperative framework could enable early warning systems for chronic disease management, elderly care, and post-operative recovery, where edge devices detect immediate abnormalities and escalate only complex or uncertain cases to a more powerful model. Such a design reduces communication overhead, protects sensitive data, and creates opportunities for personalized, always-on healthcare assistance.

%Our solution is also potential for the smart transportation and urban infrastructure. In connected traffic systems, small models deployed in vehicles, roadside units, and cameras can process local sensor streams in real time for lane monitoring, pedestrian detection, and short-term traffic prediction, while large models at regional control centers can reason over city-scale patterns, optimize signal timing, coordinate emergency response, and generate strategic mobility plans. Similar ideas apply to smart grids, industrial plants, and logistics networks, where small models ensure low-latency local control and large models provide system-wide optimization. This creates a strong opportunity for building hierarchical AI systems that are both scalable and cost-effective, combining the responsiveness of edge intelligence with the reasoning power of centralized foundation models.

However, our solution still has several limitations regarding high-efficiency robot cooperation. Explicitly, it is difficult to ensure perfect environment sensing among different robots due to limited sensing capabilities. We can extend their sensing ranges with emerging sensors, such as LiDAR and millimeter-wave sensors, through deriving reasonable path planning decisions. Nonetheless, robots struggle to select feasible neighbors to implement subsequent missions due to limited computing resources.

\section{Conclusion}
\label{sec_conclusion}
In this article, we have designed a large and small AI model cooperation framework to enable robots to achieve accurate and real-time environment sensing and estimation with feasible path planning. We have achieved deep cooperation between large AI models and small AI models to perform mutual improvement for high-efficiency sensing implementation. We proposed an adaptive model splitting and update algorithm to ensure accurate path planning and adjustment considering time-varying changes in environment. The simulation results demonstrated that our solution realized accurate environment estimation and low-latency sensing cooperation under different scenarios with various number of robots.

\bibliographystyle{IEEEtran}
\bibliography{reference}

@ARTICLE{geng,
  author={Sun, Geng and Wang, Yixian and Niyato, Dusit and Wang, Jiacheng and Wang, Xinying and Poor, H. Vincent and Letaief, Khaled B.},
  journal={IEEE Net.}, 
  title={Large Language Model (LLM)-enabled Graphs in Dynamic Networking}, 
  year={2024},
  volume={5},
  number={5},
  pages={1-1},
  doi={10.1109/MNET.2024.3511662}}

@ARTICLE{differential,
  author={Wei, Hangxing and Zhang, Gang and Du, Fuxin and Su, Jing and Wu, Jiang},
  journal={IEEE Trans. Instrum. Meas.}, 
  title={Dexterity Assessment for Wrist Joints of Surgical Robots: A Statics-Based Evaluation Method}, 
  year={2024},
  volume={73},
  number={},
  pages={1-11},
  doi={10.1109/TIM.2024.3472808}}

@ARTICLE{centralized_,
  author={Huang, Liming and Wu, Yulei and Simeonidou, Dimitra},
  journal={IEEE Transactions on Cognitive Communications and Networking}, 
  title={Optimizing System Performance for Autonomous Recovery in AI-Native Networks Using Large Language Models}, 
  year={2026},
  volume={12},
  number={},
  pages={6451-6465},
  doi={10.1109/TCCN.2026.3667165}}

@INPROCEEDINGS{FAM,
  author={Khalil, Ahmad and Wainakh, Aidmar and Zimmer, Ephraim and Parra-Arnau, Javier and Anta, Antonio Fernandez and Meuser, Tobias and Steinmetz, Ralf},
  booktitle={2023 Eighth International Conference on Fog and Mobile Edge Computing (FMEC)}, 
  title={Label-Aware Aggregation for Improved Federated Learning}, 
  year={2023},
  volume={},
  number={},
  pages={216-223},
  doi={10.1109/FMEC59375.2023.10306055}}

@ARTICLE{distributed2,
  author={Xiong, Yubing and Huang, Mingrui and Liang, Xuechen and Tao, Meiling},
  journal={IEEE Access}, 
  title={ResilioMate: A Resilient Multi-Agent Task Executing Framework for Enhancing Small Language Models}, 
  year={2025},
  volume={13},
  number={},
  pages={86892-86911},
  doi={10.1109/ACCESS.2025.3567244}}

@ARTICLE{cooperative_data,
  author={Jia, Riheng and Fu, Qiyong and Zheng, Zhonglong and Zhang, Guanglin and Li, Minglu},
  journal={IEEE/ACM Trans. Net.}, 
  title={Energy and Time Trade-Off Optimization for Multi-UAV Enabled Data Collection of IoT Devices}, 
  year={2024},
  volume={32},
  number={6},
  pages={5172-5187},
  doi={10.1109/TNET.2024.3450489}}

@article{AI_6G,
  title={Optimizing computation offloading in satellite-UAV-served 6G IoT: A deep learning approach},
  author={Mao, Bomin and Tang, Fengxiao and Kawamoto, Yuichi and Kato, Nei},
  journal={IEEE Network.},
  volume={35},
  number={4},
  pages={102--108},
  year={2021},
  publisher={IEEE}
}

@article{Aggregation,
  title={High-quality model aggregation for blockchain-based federated learning via reputation-motivated task participation},
  author={Qi, Jiahao and Lin, Feilong and Chen, Zhongyu and Tang, Changbing and Jia, Riheng and Li, Minglu},
  journal={IEEE Internet Things Journal.},
  volume={9},
  number={19},
  pages={18378--18391},
  year={2022},
  publisher={IEEE}
}

@INPROCEEDINGS{AI_driven,
  author={Picano, Benedetta and Fantacci, Romano},
  booktitle={2023 IEEE Globecom Workshops (GC Wkshps)}, 
  title={AI-Driven Digital Twins for Tasks Offloading in 6G UAV-Aided MEC Networks}, 
  year={2023},
  month = {Dec.},
  volume={},
  number={},
  pages={1093-1098},
  doi={10.1109/GCWkshps58843.2023.10464755}}

@ARTICLE{real-time_decision,  author={Mo, Yamin and Ma, Sihan and Gong, Haoran and Chen, Zhe and Zhang, Jing and Tao, Dacheng},  journal={IEEE Internet of Things Journal},   title={Terra: A Smart and Sensible Digital Twin Framework for Robust Robot Deployment in Challenging Environments},   year={2021},  volume={8},  number={18},  pages={14039-14050},  doi={10.1109/JIOT.2021.3068736}}

@ARTICLE{sensing_coverage,  author={Liu, Yi and Nie, Jiangtian and Li, Xuandi and Ahmed, Syed Hassan and Lim, Wei Yang Bryan and Miao, Chunyan},  journal={IEEE Internet of Things Journal},   title={Federated Learning in the Sky: Aerial-Ground Air Quality Sensing Framework With UAV Swarms},   year={2021},  volume={8},  number={12},  pages={9827-9837},  doi={10.1109/JIOT.2020.3021006}}

@ARTICLE{computation,  author={Huang, Qianyi and Song, Guochao and Wang, Wei and Dong, Huixin and Zhang, Jin and Zhang, Qian},  journal={IEEE Internet of Things Journal},   title={FreeScatter: Enabling Concurrent Backscatter Communication Using Antenna Arrays},   year={2020},  volume={7},  number={8},  pages={7310-7318},  doi={10.1109/JIOT.2020.2984877}}

@ARTICLE{sensing_limitation,  author={Meng, Kaitao and Wu, Qingqing and Ma, Shaodan and Chen, Wen and Quek, Tony Q. S.},  journal={IEEE Wireless Communications Letters},   title={UAV Trajectory and Beamforming Optimization for Integrated Periodic Sensing and Communication},   year={2022},  volume={11},  number={6},  pages={1211-1215},  doi={10.1109/LWC.2022.3161338}}

@ARTICLE{edge_,
  author={Li, Jiahui and Sun, Geng and Wu, Qingqing and Niyato, Dusit and Kang, Jiawen and Jamalipour, Abbas and Leung, Victor C. M.},
  journal={IEEE J. Sel. Areas Commun.}, 
  title={Collaborative Ground-Space Communications via Evolutionary Multi-Objective Deep Reinforcement Learning}, 
  year={2024},
  volume={42},
  number={12},
  pages={3395-3411},
  doi={10.1109/JSAC.2024.3459029}}

@article{collision,
  title={Deep-reinforcement-learning-based collision avoidance in uav environment},
  author={Ouahouah, Sihem and Bagaa, Miloud and Prados-Garzon, Jonathan and Taleb, Tarik},
  journal={IEEE Internet Things J.},
  volume={9},
  number={6},
  pages={4015--4030},
  year={2021},
  publisher={IEEE}
}

@INPROCEEDINGS{Convention_DDPG,  author={Chen, Weibin and Hua, Lei and Xu, Leixin and Zhang, Benshun and Li, Mengmeng and Ma, Tao and Chen, Yang-Yang},  booktitle={2021 6th ICACRE},   title={MADDPG Algorithm for Coordinated Welding of Multiple Robots},   year={2021},  volume={3},  number={5},  pages={1-5},  doi={10.1109/CACRE52464.2021.9501327}}
%\begin{IEEEbiographynophoto}
%{Longyu Zhou} is a research fellow working in the Information Systems Technology and Design Pillar at Singapore University of Technology and Design, Singapore. His research interests include Internet of Things, AI-RAN, and Digital Twins. 
%\end{IEEEbiographynophoto}
%\begin{IEEEbiographynophoto}
%{Wenjiao Feng} is currently pursuing the Ph.D. degree with the School of Information and Communication Engineering, University of Electronic Science and Technology of China. Her research interests include distributed machine learning systems and large model systems.
%\end{IEEEbiographynophoto}
%\begin{IEEEbiographynophoto}
%{Supeng Leng} is a Full Professor in the School of Information \& Communication Engineering, University of Electronic Science and Technology of China (UESTC). His research focuses on resource, spectrum, energy, routing and networking in Internet of Things, and the next generation intelligent mobile networks. 
%\end{IEEEbiographynophoto}
%\begin{IEEEbiographynophoto}
%{Mohsen Guizani} is currently a Professor of Machine Learning and the Associate Provost at Mohamed Bin Zayed University of Artificial Intelligence (MBZUAI), Abu Dhabi, UAE. His research interests include applied machine learning, artificial intelligence, Internet of Things, smart city, and cybersecurity. 
%\end{IEEEbiographynophoto}

\balance 

\end{document}